\documentclass[aps, prx, a4paper,showpacs,twocolumn,10pt,longbibliography,superscriptaddress]{revtex4-2}
\usepackage{bbm, amsmath, amssymb, amsthm, bm,textcomp, nicefrac,geometry,ragged2e}
\usepackage{ dsfont }
\usepackage{graphicx,epstopdf,color,verbatim,enumitem,ulem,etoolbox}
\usepackage[dvipsnames]{xcolor}
\usepackage[bbgreekl]{mathbbol}
\usepackage[unicode=true,pdfusetitle, bookmarks=false,bookmarksnumbered=false,
bookmarksopen=false, breaklinks=false,pdfborder={0 0 0},backref=false,
colorlinks=true, linkcolor=myrefcolor,citecolor=myurlcolor,urlcolor=myurlcolor]{hyperref}
\definecolor{myurlcolor}{rgb}{0.05,0.45,0.2}
\definecolor{myrefcolor}{rgb}{0.9,0.1,0.45}
\usepackage{pbox,array}
\usepackage{algcompatible}
\usepackage{algpseudocode}
\usepackage[ruled,vlined]{algorithm2e}
\usepackage{diagbox}
\usepackage{tikz}

\makeatletter
\appto\appendix{%
  \@ifstar{\def\theequation@prefix{A.}}%
          {}%
}
\makeatother

\DeclareSymbolFontAlphabet{\mathbb}{AMSb}
\DeclareSymbolFontAlphabet{\mathbbl}{bbold}

\newcommand{\bra}[1]{\left\langle #1\right|}

\newcommand{\ket}[1]{\left|#1\right\rangle}

\newcommand\cH{{\mathcal H}}

\newcommand\cF{{\mathcal F}}

\newcommand\cB{{\mathcal B}}

\newcommand\cO{{\mathcal O}}

\newcommand\id{\mathbbl{1}}

\newcommand*\colvec[1]{\begin{pmatrix}#1\end{pmatrix}}

\definecolor{darkspringgreen}{rgb}{0.09, 0.45, 0.27}

\newcommand{\be}{\begin{equation}}
\newcommand{\ee}{\end{equation}}
\newcommand{\bea}{\begin{eqnarray}}
\newcommand{\eea}{\end{eqnarray}}

\newcommand{\tab}[1]{Tab.~\ref{tab:#1}}
\newcommand{\fig}[1]{Fig.~\ref{#1}}
\newcommand{\alg}[1]{Alg.~\ref{#1}}
\newcommand{\eq}[1]{Eq.~(\ref{eq:#1})}

\SetKw{KwBy}{by}

\def\tr{\mathrm{tr}}

\begin{document}

\title{Self-testing two-qubit maximally entangled states from generalized CHSH tests}
\author{Xavier Valcarce}
\thanks{both authors contribute equally to this work.}
\affiliation{Universit\'e Paris-Saclay, CEA, CNRS, Institut de physique th\'eorique, 91191, Gif-sur-Yvette, France}
\affiliation{Departement Physik, Universit\"at Basel, Klingelbergstrasse 82, 4056 Basel, Switzerland}
\author{Julian Zivy}
\thanks{both authors contribute equally to this work.}
\affiliation{Universit\'e Paris-Saclay, CEA, CNRS, Institut de physique th\'eorique, 91191, Gif-sur-Yvette, France}
\affiliation{Departement Physik, Universit\"at Basel, Klingelbergstrasse 82, 4056 Basel, Switzerland}
\author{Nicolas Sangouard}
\affiliation{Universit\'e Paris-Saclay, CEA, CNRS, Institut de physique th\'eorique, 91191, Gif-sur-Yvette, France}
\affiliation{Departement Physik, Universit\"at Basel, Klingelbergstrasse 82, 4056 Basel, Switzerland} 
\author{Pavel Sekatski}
\affiliation{Departement Physik, Universit\"at Basel, Klingelbergstrasse 82, 4056 Basel, Switzerland}
\begin{abstract}
Device-independent certification, also known as self-testing, aims at guaranteeing the proper functioning of untrusted and uncharacterized devices. For example, the quality of an unknown source expected to produce two-qubit maximally entangled states can be evaluated in a bi-partite scenario, each party using two binary measurements. The most robust approach consists in deducing the fidelity of produced states with respect to a two-qubit maximally entangled state from the violation of the CHSH inequality.  In this paper, we show how the self-testing of two-qubit maximally entangled states is improved by a refined analysis of measurement statistics. The use of suitably chosen Bell tests, depending on the observed correlations, allows one to conclude higher fidelities than ones previously known. In particular, nontrivial self-testing statements can be obtained from correlations that cannot be exploited by a CHSH-based self-testing strategy. Our results not only provide novel insight into the set of quantum correlations suited for self-testing, but also facilitate the experimental implementations of device-independent certifications.  
\end{abstract}

\maketitle 

\section{Introduction}
Bell inequalities were proposed to show that the results of local incompatible measurements on subsystems prepared in a global quantum state can have stronger-than-classical correlations, so-called non-local correlations~\cite{Bell1964}. Self-testing aims to reconstruct the global state and the local measurements generating these non-local correlations (up to local isometries) in a device-independent way, i.e. without assumption on the functioning of devices used in the Bell test~\cite{Montanaro2016,Supic2019}.\bigskip


A self-test of a two-qubit maximally entangled state $\phi^+_{AB}$ for example can be obtained from the simplest Bell inequality -- the Clauser-Horne-Shimony-Holt (CHSH) inequality~\cite{Clauser1969}. The latter is tested in a bi-partite scenario in which Alice and Bob share a state $\rho_{AB}\in \cH_A \otimes \cH_B$ and perform one out of two binary measurements each. The measurement choice is labelled $A_x,B_y$ where $x,y =0,1$ for Alice and Bob, respectively, with the results for each measurement choice $\pm 1.$ By repeating the experiment many times, the \textit{CHSH score} is computed from 
\begin{equation}
    \beta = \langle A_0(B_0+B_1) \rangle + \langle A_1(B_0-B_1) \rangle\,,
\end{equation}
where $\langle A_x B_y \rangle$ is the expectation values of results for the measurement choice $A_x$ and $B_y$ (the probability that the results are the same minus the probability that they are different). The CHSH \textit{inequality} is obtained by noting that, for any locally causal theory, $\beta$ is upper bounded by $2$ -- the local bound~\cite{Brunner14}. Hence, the observation of any score $\beta>2$ rules out the possibility to model the experiment in the framework of such theories. Furthermore, the sole knowledge of the CHSH score can be used to certify that the shared state $\rho_{AB}$ resembles $\phi^+_{AB}$~\cite{Popescu92, Braunstein92, Mayers2004}. More precisely, when $\beta$ is sufficiently close to $2\sqrt{2}$, there exist local maps $\Lambda_A$ and $\Lambda_B$ that can be applied to $\rho_{AB}$ to extract a two-qubit state $\Lambda_A \otimes \Lambda_B[\rho_{AB}]$ having a fidelity with respect to $\phi^+_{AB}$ which exceeds  $1/2$~\cite{Bardyn2009,Bancal15,Yang2014, Kaniewski2016}.\bigskip

Self-testing by the CHSH score is not only an elegant characterisation of bipartite sources, but has been shown to be a key ingredient for the device-independent certification of other quantum instruments, including quantum processing units~\cite{Magniez2006, Sekatski2018} and generalised measurements~\cite{Wagner2020}. Its implementation is however not easy. Self-testing is simply not possible for CHSH scores smaller than $\beta \approx 2.05$~\cite{Valcarce2020} and a non-trivial fidelity with respect to two-qubit maximally entangled states can only be extracted for a CHSH score exceeding $\beta \gtrapprox 2.11$~\cite{Kaniewski2016}. This explains why first experimental realisations~\cite{Wang2018,Goh2019,Gomez2019} were not device-independent~\cite{Orsucci2020} and only one self-testing realisation~\cite{Bancal2021} has so far been reported. In order to popularize self-testing, it is natural to ask if the noise tolerance for the self-testing of $\phi^+_{AB}$ can be improved. \bigskip 



To achieve this goal, an approach consists in deriving self-tests from a more refined analysis of measurement statistics. Indeed, in order to evaluate the CHSH score experimentally, all four expectation values $\langle A_x B_y \rangle$ have to be collected. These values give more information than just the CHSH score, and can be used to generate all the self-testings of the singlet~\cite{Wang2016}. Note that first attempts for robust self-testing were based on the knowledge of these individual expectation values but none of them succeeded to provide non-trivial state fidelities with respect to two-qubit maximally entangled states if the CHSH score $\beta < 2.37$~\cite{Bancal2015}.\bigskip

Here we focus on a family of Bell tests with a \textit{generalized CHSH score} given by
\begin{equation}
    \label{eq:btheta}
   \beta_{\theta} = \sqrt{2}(\cos(\theta)\underbrace{\langle A_0(B_0+B_1)\rangle}_{X} + \sin(\theta)\underbrace{\langle A_1(B_0-B_1)\rangle}_{Y} ),
\end{equation}
for some $\theta\in[0,\frac{\pi}{2}]$. The local bound is given by $\beta_\theta^L=2\sqrt{2}\max(\cos{\theta},\sin{\theta})$~\cite{Acin2012, Woodhead2020, Sekatski2020} and the Bell inequalities $\beta_\theta < \beta_\theta^L$ can be tested in a CHSH scenario, where two parties dispose of two binary measurements each. Note that the case $\theta=\pi/4$ reduces to the CHSH case. We show that this family of generalized CHSH tests can be used to self-test two-qubit maximally entangled states with a fidelity higher than with the CHSH score whenever $X \neq Y.$ In some cases, a non-trivial fidelity with respect to the two-qubit maximally entangled state can be extracted from a suitably chosen generalized CHSH test, even when the CHSH score $\beta$ is larger than, but arbitrarily close to, $2.$ We conclude with an explicit recipe to choose the test giving the highest fidelity in any experiments where the values of correlators $X$ and $Y$ are measured. 

\section{Preliminaries on self-testing}
We stay with the scenario described in the introduction. Two protagonists, Alice and Bob, share an unknown state $\rho_{AB}\in \cH_A \otimes \cH_B$ with unknown local Hilbert space dimensions and perform one out of two binary measurements each, $A_x$ and $B_y$ with $x,y = 0,1$ denoting the respective measurements of Alice and Bob. By repeating the experiment many times, Alice and Bob obtain the individual expectation values X and Y defined in Eq.~\eqref{eq:btheta}.
From these expectation values, they want to show that the state $\rho_{AB}$ resembles a maximally-entangled two-qubit state $\phi_{AB}
^+.$ In the framework of self-testing, we express this resemblance by a notion of \textit{extractability} or \textit{singlet fraction}, defined by
\begin{equation}
    \label{eq:extr}
    \Xi[\rho_{AB}\!\rightarrow\phi^+_{AB}]=\max_{\Lambda_A, \Lambda_B} F( (\Lambda_A \otimes \Lambda_B) [\rho_{AB}], \phi^+_{AB})\,.
\end{equation}
The former function captures the maximal fidelity with the target state $\phi^+_{AB}$ for a given state $\rho_{AB}$ over all local completely positive trace preserving (CPTP) maps $\Lambda_A \otimes \Lambda_B$~\cite{Bardyn2009}. The local maps $\Lambda_{A/B}: \mathcal{L}(\cH_{A/B}) \rightarrow \mathcal{L}(C^2)$ should be thought of as necessary operations to identify a degree of freedom of the whole system $\rho_{AB}$ that can be described by a two-qubit state. The fidelity is the square of the Uhlmann fidelity $F(\rho_0,\rho_1)=\left(\tr\left[\sqrt{\rho_0^{1/2}\rho_1 \rho_0^{1/2}}\right]\right)^2$ which reduces to the overlap $F(\rho_0,\rho_1)=\tr\left[\rho_0 \rho_1 \right]$ whenever one of the two states $\rho_{0/1}$ is pure. We will simply denote the extractability of $\rho_{AB}$ by $\Xi[\rho_{AB}]$ in the rest of the manuscript as there will be no ambiguity to the reference state. \bigskip

We are interested in bounding the extractability $\Xi[\rho_{AB}]$ as a function of observed quantities X and Y. Formally, this is accomplished by considering all possible quantum models $(\rho_{AB},A_x,B_y)$ satisfying $\tr\left(\rho_{AB} \left(A_0\left(B_0 + B_1\right)\right)\right) \geq X$ and $\tr\left(\rho_{AB} \left(A_0\left(B_0 - B_1\right)\right)\right) \geq Y.$ However, the set of correlator-pairs $(X,Y)$, for which the extractability is lower-bounded by a constant, is convex: two quantum models can be joined into a new model on which the extractability is bounded by the weighted sum of extractabilities associated to the individual models. It is thus equivalent to bound the extractability from linear constraints of $X$ and $Y$, i.e. from the Bell score $\beta_\theta$. Our aim is thus to solve the following optimization
\begin{eqnarray}
\label{fidelity}
&&\cF=\min_{\rho_{AB}, A_x, B_y}\Xi[\rho_{AB}]\\
\nonumber
&&
\text{s.t.} \, \tr \left(\rho_{AB} \mathcal{B}_{\theta}\right) \geq \beta_\theta,
\end{eqnarray}
where 
\begin{equation}
\mathcal{B}_\theta=\sqrt{2}(\cos(\theta) A_0(B_0+B_1) + \sin(\theta) A_1(B_0-B_1))
\end{equation}
is a Bell operator identified as the \textit{generalized CHSH operator}. This is a hard optimization problem given that the dimensions of the Hilbert spaces supporting the initial states are unknown and that the set of product maps is non-convex. Before approaching the problem \eqref{fidelity}, we note that there exists a trivial strategy to achieve $\Xi[\rho_{AB}]=\frac{1}{2}$ for all state $\rho_{AB}$. Indeed, it is always possible for Alice and Bob to choose CPTP maps destroying the shared state and replacing it by a fixed state of their choice, resulting in a product state $\rho_{AB} = \rho_A \otimes \rho_B$. Alice and Bob can always chose a product state with a fidelity of $\frac{1}{2}$ with respect to a two-qubit maximally entangled state. We thus call this fidelity the \textit{trivial fidelity}.

\section{Self-testing with two binary measurements}
For the particular case of interest where two dichotomic measurements are used by each party, we can choose a basis according to Jordan's lemma so that the local observables $A_x$ and $B_y$ are block diagonal with blocks of size $2\times 2$ represented by Pauli measurements
\begin{align}
\begin{split}
    A_x = \bigoplus_i A_x^i &= \bigoplus_i \cos{(a_i)}\sigma_h + (-1)^x \sin{(a_i)}\sigma_m, \\
    B_y = \bigoplus_j B_y^j &= \bigoplus_j \cos{(b_j)}\sigma_z + (-1)^y \sin{(b_j)}\sigma_x.
\end{split}
\end{align}
$\sigma_{h,m}=\frac{1}{\sqrt{2}}(\sigma_z \pm \sigma_x)$ where $\sigma_x$ and $\sigma_z$ are the Pauli operators and $a_i,b_j \in [0,\frac{\pi}{2}]$ are the angles between the local measurements in the blocks defined by the indexes $i$ and $j.$ Trivially, our family of Bell operators inherits the same block structure 
\begin{equation}
    \cB_\theta = \bigoplus_{ij} \cB_\theta^{i,j}.
\end{equation}
A priori, the state $\rho_{AB}$ does not have the same diagonal structure. However, Alice and Bob can locally perform a projection into their orthogonal blocks before choosing the extraction map. Let us call $p_{ij}$ the probability to get a successful projection into the blocks $i$ for Alice and $j$ for Bob, and $\rho_{AB}^{ij}$ the resulting state. At first sight, the extractability of $\rho_{AB}$ is larger than $\sum_{ij} p_{ij} \rho_{AB}^{ij}$, but because of the block diagonal structure of the measurements, the two states lead to the same score $\beta_\theta$ and hence have the same extractability. Without loss of generality, we can thus consider that $\rho_{AB}$ is of the form 
\begin{equation}
\rho_{AB}=\sum_{ij} p_{ij} \rho_{AB}^{ij}
\end{equation}
This means that the extraction maps can be constructed independently for each block
\begin{equation}
\Lambda_{A}^i: \mathcal{L}(C^2) \rightarrow \mathcal{L}(C^2), \, \-\  \Lambda_{B}^j: \mathcal{L}(C^2) \rightarrow \mathcal{L}(C^2)
\end{equation}
and that the extractability \eqref{eq:extr} reduces to a maximization over qubit maps 
\begin{equation}
    \label{eq:extrqubit}
    \Xi[\rho_{AB}]=\max_{\Lambda_A^i, \Lambda_B^j} \sum_{ij} p_{ij} F((\Lambda_A^i \otimes \Lambda_B^j) [\rho_{AB}^{ij}], \phi^+_{AB})\,.
\end{equation}
If we fix the dependence of the extraction maps on the angles $a_i$ and $b_j$ and consider the case where they do not depend on the state,  we can lower bound the extractability $\cF$ given in Eq.~\eqref{fidelity} by first solving the following optimization
\begin{eqnarray}
\label{eq:O}
&\mathcal{O}_{\min}(\beta_\theta')& =\min_{a_i,b_j,\rho_{AB}^{\text{qubit}}} F(\Lambda_A^i \otimes \Lambda_B^j[\rho_{AB}^{\text{qubit}}],\phi_{AB}^+)\\
\nonumber
&& \text{s.t.} \quad \tr(\cB_\theta^{i,j} \rho_{AB}^{\text{qubit}}) \geq \beta_\theta'
\end{eqnarray}
for all physical two-qubit states $\rho_{AB}^{\text{qubit}}$ and then taking the convex roof of $\mathcal{O}_{\min}(\beta_\theta')$. The optimization over states for fixed measurement angles is a minimization of a linear objective function over a spectrahedron. Such problems can  be  efficiently  solved using semi-definite programming. The minimization over angles between  measurements  can  be  done  using  a  non-linear optimisation algorithm. The details for these optimisations are detailed in the Appendix. When considering self-testing from the CHSH operator $\mathcal{B}_{\theta=\frac{\pi}{4}},$ the result of this two-fold optimisation followed by a convex roof shows that a non-trivial fidelity can be obtained as long as $\beta \gtrapprox 2.11.$ This takes a clever choice of local maps that we present in the following section. 

\section{Choice of maps for self-testing from the CHSH operator}
As stated in the previous section, local maps need to be fixed to bound the self-testing fidelity from the numerical optimization given in \eq{O}. A particularly relevant choice has been reported by Kaniewski in Ref. \cite{Kaniewski2017} for self-testing a maximally entangled two-qubit state from the CHSH operator. We quickly present the basic ideas leading to this choice of maps and, in the next section, show how to use the same line of thought to construct maps relevant for self-testing a maximally entangled two-qubit state from the generalized CHSH operator $\mathcal{B}_\theta.$ \bigskip

Consider the CHSH operator appearing in the constraint of \eq{O} for the block characterized by the index $i$ and $j$ for Alice and Bob, respectively. If we forget these indices, the CHSH operator takes the following form
\begin{equation}
\cB_{\pi/4}= 2\, \sum_{k,\ell=0,1} M_{k,\ell}(a,b)\,  \widehat{\sigma}_k \otimes \overline{\sigma}_\ell
\end{equation}
with $M_{k,\ell}(a,b)$ the elements of the matrix
\begin{equation}
\nonumber
M(a,b) = \left(\begin{array}{cc}
\cos(a) \cos(b) & \cos(a) \sin(b) \\
\sin(a) \cos(b) & -\sin(a) \sin(b)
\end{array}\right)
\end{equation}
and $\{\widehat{\sigma}_0, \widehat{\sigma}_1\}=\{\sigma_h,\sigma_m\}$ and $\{\overline{\sigma}_0, \overline{\sigma}_1\}=\{\sigma_z,\sigma_x\}$. The eigenvalues $\lambda_i(a,b),$ $i={1,\dots,4}$ with $\lambda_1 \geq \lambda_2 \geq \lambda_3 \geq \lambda_4$ of this single-block CHSH operator can be written in the decreasing order as
\be\begin{split}
     {\bm \lambda}(a,b)&=\text{Eig}^\downarrow(\cB_{\pi/4}) \\
     \nonumber
&=\left(\begin{array}{c}
\sqrt{2}\sqrt{2 + \cos\big(2(a-b)\big)- \cos\big(2(a+b)\big)}\\
\sqrt{2}\sqrt{2 - \cos\big(2(a-b)\big)+ \cos\big(2(a+b)\big) }\\
-\sqrt{2}\sqrt{2 - \cos\big(2(a-b)\big)+ \cos\big(2(a+b)\big)}\\
-\sqrt{2}\sqrt{2 + \cos\big(2(a-b)\big)- \cos\big(2(a+b)\big)}
\end{array}\right).
\end{split}
\ee
We denote the corresponding eigenstates $\{ \ket{\psi_i(a,b)}\}_{i=1,\dots,4}$. For fixed measurement angles, the Bell score $\beta'_{\frac{\pi}{4}}$ of any two-qubit state $\rho_{AB}^{\text{qubit}}$ is therefore given by
\be
\beta'_{\frac{\pi}{4}} = \sum_{i=1}^4  \lambda_i(a,b) \bra{\psi_i(a,b)} \rho_{AB}^{\text{qubit}}\ket{\psi_i(a,b)}.
\ee
On one hand, the maximal eigenvalue satisfies $\lambda_1(a,b) \in[2,2\sqrt{2}]$, while $\lambda_2(a,b) \in[0,2].$ On the other hand, the two lowest eigenvalues $\lambda_3$ and $\lambda_4$ are negative, and the states with support on $\ket{\psi_3(a,b)}$ and $\ket{\psi_4(a,b)}$ thus lead to a relatively low CHSH score. We observed numerically that an appropriate design of local maps can be obtained by considering states supported on the two-dimensional subspace $\widehat{\cH}_{\pi/4}(a,b)$ spanned by the first two eigenstates. These maps are obtained by considering various values of parameters $a$ and $b.$ \bigskip

The maximal CHSH score corresponding to $\lambda_1=2\sqrt{2}$ can only be attained for $a=b=\frac{\pi}{4}$ (in which case $\lambda_2=0$). The corresponding state $\ket{\psi_1(\frac{\pi}{4},\frac{\pi}{4})}$ is equal to the target state $\phi^+_{AB}$. Choosing the extraction maps $\Lambda_A(\frac{\pi}{4}) \otimes \Lambda_B(\frac{\pi}{4})$ that leave the state $\phi^+_{AB}$ unchanged guarantees that $\mathcal{O}_{\min}(\beta_{\frac{\pi}{4}}'=2\sqrt{2})=1.$ We thus set local maps for $(a,b)=(\frac{\pi}{4},\frac{\pi}{4})$ as the identity.\bigskip

Let us now consider the values $a,b\in\{ 0,\frac{\pi}{4},\frac{\pi}{2}\}$, where the Bell operator takes the form
\begin{center}
    \begin{tabular}{ c| c c c }
       \multicolumn{4}{c}{$\frac{1}{2} \cB_{\pi/4}(a,b)$} \\
    \hline
        \diagbox{b}{a}  & $0$ & $\frac{\pi}{4}$ & $\frac{\pi}{2}$ \\
    \hline
    $0$ & $\sigma_h\otimes\sigma_z$ & $\sigma_z\otimes\sigma_z$ & $ \sigma_m\otimes\sigma_z $ \\
    $\frac{\pi}{4}$ & $\sigma_h\otimes\sigma_h$ & &  $\sigma_m\otimes\sigma_m$ \\  
    $\frac{\pi}{2}$ & $\sigma_h\otimes\sigma_x$ & $\sigma_x\otimes\sigma_x$ & $-\sigma_m\otimes\sigma_x$   
\end{tabular}
\end{center}
When either $a=\frac{\pi}{4}$ or $b =\frac{\pi}{4},$ the CHSH operator takes the form $\cB_{\pi/4} = 2\, \sigma_{\bf n}\otimes \sigma_{\bf n}$ with the Bloch vector $\bf n$ in the X-Z plane. Consequently, the two maximal eigenvalues are degenerate, $\lambda_1=\lambda_2=2$, and the corresponding subspace is spanned by the states
\be\begin{split}
\phi_{AB}^+=\frac{1}{4}(\id\otimes\id + \sigma_{\bf n} \otimes \sigma_{\bf n} + \sigma_{\bf n_\perp}\otimes \sigma_{\bf n_\perp} - \sigma_y\otimes\sigma_y)\\
\phi_{AB}^\perp =\frac{1}{4}(\id\otimes\id + \sigma_{\bf n} \otimes \sigma_{\bf n} - \sigma_{\bf n_\perp}\otimes \sigma_{\bf n_\perp} + \sigma_y\otimes\sigma_y)\nonumber\end{split}
\ee
where ${\bf n_\perp}$ is the Bloch vector orthogonal to ${\bf n}$ in the Z-X plane.  Hence, in order to attain a non-zero extraction fidelity at $\beta_{\frac{\pi}{4}}'=2$, the maps have to deform the Bloch sphere associated to the two dimensional space. Let us keep in mind that for the angle $a=\frac{\pi}{4},$ the extraction map is given by the identity (and similarly for $b=\frac{\pi}{4}$). The other party thus has to select a map that increases the overlap between the state $\phi_{AB}^\perp$ and the target state $\phi^+_{AB}$ (they are manifestly orthogonal to start with). This can be done by choosing a dephasing map in the direction $\sigma_{\bf n},$ that is, for angles $0$ and $\frac{\pi}{2},$ the map changes $\rho$ into $\frac{1}{2}\left(\rho + \sigma_{\bf n} \, \rho \, \sigma_{\bf n} \right).$ Since any state in the subspace spanned by  $\ket{\phi_{AB}^+}$ and $\ket{\phi_{AB}^\perp}$ can be expressed as
 \be
\rho_{AB} =\frac{1}{2}(\Sigma_0 + {\bf r}\cdot {\bm \Sigma})
 \ee
with a unit vectors $\bf r$, and the operators
\be\begin{split}
    \Sigma_0 &= \frac{1}{2} (\id\otimes \id + \sigma_{\bf n}\otimes \sigma_{\bf n})\\
    \Sigma_1 &= \frac{1}{2} ( \id \otimes \sigma_{\bf n}+ \sigma_{\bf n}\otimes \id )\\
    \Sigma_2 &= \frac{1}{2} (\sigma_{\bf n_\perp}\! \otimes \sigma_{\bf n_\perp} -  \sigma_{y}\otimes \sigma_{y})\\
    \Sigma_3 &= -\frac{1}{2} (\sigma_{\bf n_\perp}\! \otimes \sigma_{y} +  \sigma_{y}\otimes \sigma_{\bf n_\perp}),
\end{split}
\ee
the map changing $\rho$ into $\frac{1}{2}\left(\rho + \sigma_{\bf n} \, \rho \, \sigma_{\bf n} \right)$
sends any state in the subspace of interest onto
\be\label{eq: state final}
\frac{1}{4}(\id\otimes\id + \sigma_{\bf n} \otimes \sigma_{\bf n}) + \frac{x}{4}(\id \otimes \sigma_{\bf n} +  \sigma_{\bf n} \otimes \id)
\ee
 with $|x|\leq 1.$ Such a state has a fidelity with the target state $\phi_{AB}^+$ equal to $1/2$. \bigskip
 
It is interesting to consider what these maps do for the remaining points in the table above. It turns out that the situation is similar for all the remaining combinations of angles and can be illustrated by considering the example $(a,b)=(0,0)$ only. In this case, the maps on both Alice's and Bob's side are dephasing maps in the directions $\sigma_h$ and $\sigma_z$ respectively. Any state attaining $\beta=2$ is sent by these maps to states of the form given by Eq.~\eqref{eq: state final}, where $\sigma_{\bf n}$ is replaced by $\sigma_h$ for Alice and $\sigma_z$ for Bob. It is easy to see that the fidelity of such states with the target state is given by
\be
F_L = \frac{2+\sqrt{2}}{8}\approx 0.43.
\ee
Remarkably, if  we connect the points $(2,F_L)$ and $(2\sqrt{2},1)$ in the $(\beta,F)$ plane, we see that it intersects the trivial fidelity line $F=1/2$ at precisely $\beta_{\frac{\pi}{4}}^t=\frac{2(8+7\sqrt{2})}{17} \approx 2.11$. We thus succeeded to identify the angles and states that tightly constrain the final extractability bound.\bigskip

Finally, to get the extraction maps for all angles $a,b \in [0,\frac{\pi}{2}]$, one simply does an analytical continuation between $\Lambda_A(0)\longleftrightarrow \Lambda_A(\frac{\pi}{4}) \longleftrightarrow\Lambda_A(\frac{\pi}{2})$ and similarly for $\Lambda_B$. In particular, we  take the one proposed by Kaniewski~\cite{Jed2016}
\begin{equation}
    \label{eq:mapsjed}
    \Lambda_{A}(a) = \left( \frac{1+g(a)}{2}\id\rho\id+\frac{1-g(a)}{2}\Gamma(a)\rho\,\Gamma(a)\right),
\end{equation}
with the strength 
\be
\label{strength_maps}
g(a)=(1+\sqrt{2})(\cos{a}+\sin{a}-1)
\ee
and the dephasing direction
\begin{equation}
    \Gamma(a) = \begin{cases}
      \sigma_h, & \text{if}\ a\leq\frac{\pi}{4} \\
      \sigma_m, & \text{otherwise}.
    \end{cases} \\
\end{equation}
Bob's map is similar to the one of Alice, but with a dephasing in the direction
\begin{equation}
    \Gamma(b) = \begin{cases}
      \sigma_z, & \text{if}\ b\leq\frac{\pi}{4} \\
      \sigma_x, & \text{otherwise}.
    \end{cases} \\
\end{equation}
Interestingly, numerical results suggest that the exact form of the function $g$ on intermediate angles is not important for the final bound found in Ref.~\cite{Jed2016}.

\section{Extraction maps for self-testing from generalized CHSH operators}

We now consider the eigenvalues of generalized Bell operators
\be
{\bm \lambda}^{(\theta)}(a,b) = \text{Eig}^\downarrow(\cB_{\theta}(a,b)).
\ee
The last two eigenvalues $\lambda_4^{(\theta)} =-\lambda_1^{(\theta)}$ and $\lambda_3^{(\theta)}=-\lambda_2^{(\theta)}$ are still negative, and as before, they are not considered in the construction of relevant maps. Hence, we can start by constructing extraction maps using reasoning analogous to the one presented in the previous section. \bigskip

Since the maximal quantum value $\lambda_1^{(\theta)}(\frac{\pi}{4},\theta) =2\sqrt{2}$ of the Bell operator $\cB_{\theta}(a,b)$ is attained for the setting choice $(a,b)=(\frac{\pi}{4},\theta),$ we choose maps satisfying $\Lambda_A(\frac{\pi}{4})=\Lambda_B(\theta)=\id$, so that the target state is extracted when the Bell score is maximal. \bigskip

We now consider the value of Bell operators $\cB_{\theta}(a,b)$ on the frame $(a,b)\in \partial([0,\frac{\pi}{2}]\times[0,\frac{\pi}{2}])$. It is still given by products of Paulis, with the important difference that the prefactors now depend on the angle $b$ as shown in the following table
\begin{table}[ht]
    \centering
    \small
    \begin{tabular}{ c| c c c }
        \hline
        \multicolumn{4}{c}{$\frac{1}{2}\cB_{\theta}(a,b)$} \\
    \hline
        \diagbox{b}{a}  & $0$ & $\ldots$ & $\frac{\pi}{2}$ \\
    \hline
    $0$ & $f(0)\sigma_h\otimes\sigma_z$ & $f(0)\,A_0(a)\otimes \sigma_z$ & $f(0) \sigma_m\otimes\sigma_z$ \\
    $\vdots$ & $f(b)\,\sigma_h\otimes \sigma^{(+)}(b) $ & &  $f(b)\,\sigma_m\otimes \sigma^{(-)}(b)$ \\  
    $\frac{\pi}{2}$ & $f(\frac{\pi}{2})\sigma_h\otimes\sigma_x$ & $f(\frac{\pi}{2})\, A_1(a)\otimes \sigma_x$ & $-f(\frac{\pi}{2}) \sigma_m\otimes\sigma_x$   
    \end{tabular}
    \caption{The Bell operator $\frac{1}{2}\cB_{\theta}(a,b)$ on the frame $(a,b)\in \partial([0,\frac{\pi}{2}]\times[0,\frac{\pi}{2}])$ takes the form of a tensor product of local Pauli operators.}
    \label{tab:frame}
\end{table}

Here, $f(b) =\sqrt{1+\cos(2b)\cos(2 \theta)}$ is a monotonic function ranging from $f(0)=\sqrt{2}\cos(\theta)$  to $f(\frac{\pi}{2}) = \sqrt{2}\sin(\theta)$ and 
\be
    \sigma^{(\pm)}(b)= \frac{\cos(b)\cos(\theta)\sigma_z \pm \sin(b)\sin(\theta)\sigma_x}{\sqrt{\cos^2(b)\cos^2(\theta)+\sin^2(b)\sin^2(\theta)}}.
\ee

\begin{figure}
\includegraphics[width=.47\textwidth]{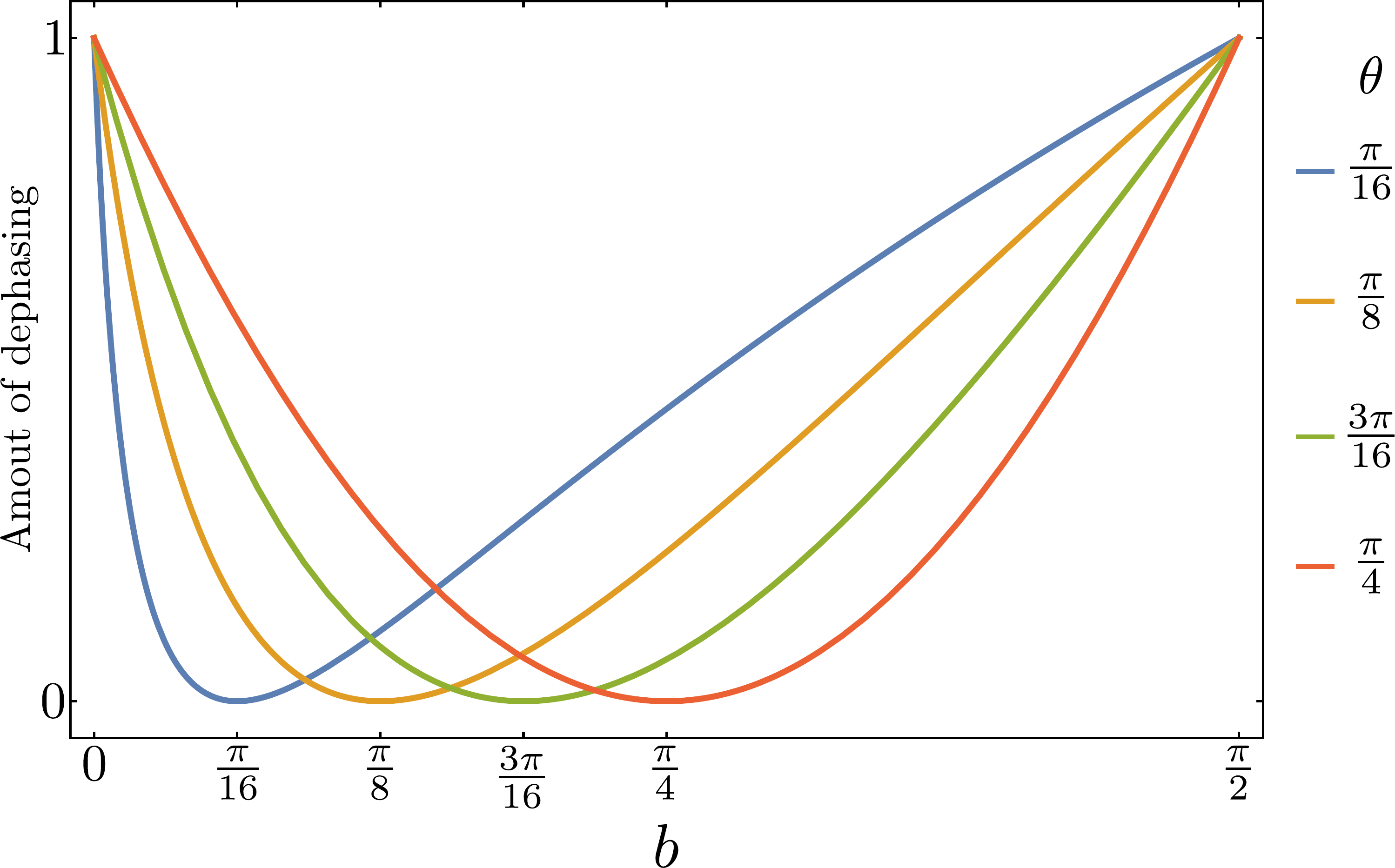}
\caption{Strength of dephasing $\tilde{g}_\theta(b)$ of Bob's local extraction maps as a function of the angle of measurement $b$ for different Bell operators $\cB_\theta$ characterized by  $\theta=\{\frac{\pi}{16},\frac{\pi}{8},\frac{3\pi}{16},\frac{\pi}{4}\}$.}
\label{fig:dephasing}
\end{figure}

As the Bell operator $\cB_{\theta}(a,b)$ takes the same value as the CHSH operator $\cB_{\frac{\pi}{4}}(a,b)$ for the extreme angles $a,b=0,\frac{\pi}{2},$ it is still relevant to align the dephasing direction of maps with the direction of the corresponding Pauli operators $\sigma_h, \sigma_m,\sigma_z$ or $\sigma_x$. We thus temporarily fix $\Lambda_{A(B)}$ to be of the same form as in Eq.~\eqref{eq:mapsjed}. To fix the strength of the dephasing for intermediate angles, we again resort to an analytic continuation. For Alice, we pick the strength as in Eq.~\eqref{strength_maps}. For Bob, the situation is different as the continuation has to be done between the angles $0\longleftrightarrow \theta \longleftrightarrow  \frac{\pi}{2}$ where the maps are fixed. Inspired by Ref.~\cite{Wagner2020}, we define a new function
\begin{equation}
    \tilde{g}_\theta(b) =(1+\sqrt{2})(\cos(t(b,\theta))+\sin(t(b,\theta))-1)
\end{equation}
where
\begin{align}
    t(b,\theta) & = \begin{cases}
        b, & \text{if}\ \theta=\frac{\pi}{4} \\
        \gamma^{-1}\ln{\frac{b+\delta_\theta}{\delta_\theta}}, & \text{otherwise}
    \end{cases} \\
    \gamma & = \frac{4}{\pi}\ln\left(\frac{\frac{\pi}{2}-\theta}{\theta}\right)\,, \\
    \delta_\theta & = \frac{\theta^2}{\frac{\pi}{2}-2\theta}\ .
\end{align}
The function $\tilde{g}_\theta(b),$ i.e. the variation of the strength of dephasing, is plotted in \fig{fig:dephasing} as a function of the angle $b$ for different choice of Bell operators ($\theta$). \bigskip

On the frame, i.e. $(a,b)\in \partial([0,\frac{\pi}{2}]\times[0,\frac{\pi}{2}]),$ the maps we just defined transform the states associated to positive eigenvalues of the Bell operators in states of the form given in Eq.~\eqref{eq: state final} with two different indexes for Alice and Bob. Specifically, $\sigma_\mathbf{n} \rightarrow \sigma_A(a)$ for Alice and $\sigma_\mathbf{n} \rightarrow \sigma_B(b)$ for Bob with 
\be
\begin{split}(\sigma_A(a),\sigma_B(b))= \nonumber
  \begin{cases}
    (c_a \sigma_h + g(a)s_a \sigma_m,\sigma_z) & b=0,a\leq \frac{\pi}{4}\\
    (g(a) c_a \sigma_h + s_a \sigma_m,\sigma_z) & b=0, a> \frac{\pi}{4}\\
    (c_a \sigma_h - g(a) s_a \sigma_m,\sigma_x) & b=\frac{\pi}{2},a\leq \frac{\pi}{4}\\
    (g(a) c_a \sigma_h - s_a \sigma_m,\sigma_x) &b=\frac{\pi}{2},a> \frac{\pi}{4}\\
    (\sigma_h,\frac{c_b c_\theta \sigma_z+\tilde{g}(b)s_b s_\theta \sigma_x}{\sqrt{(c_b c_\theta)^2+(s_b s_\theta})^2})& a=0,b\leq \theta\\
    (\sigma_h,\frac{\tilde{g}_\theta(b) c_b c_\theta \sigma_z+s_b s_\theta \sigma_x}{\sqrt{(c_b c_\theta)^2+(s_b s_\theta})^2})& a=0,b> \theta\\
    (\sigma_m,\frac{c_b c_\theta \sigma_z-\tilde{g}_\theta(b)s_b s_\theta \sigma_x}{\sqrt{(c_b c_\theta)^2+(s_b s_\theta})^2})& a=\frac{\pi}{2},b\leq \theta\\
    (\sigma_m,\frac{\tilde{g}_\theta(b) c_b c_\theta \sigma_z-s_b s_\theta \sigma_x}{\sqrt{(c_b c_\theta)^2+(s_b s_\theta})^2})& a=\frac{\pi}{2},b> \theta\\
  \end{cases}
\end{split}
\ee
The fidelity of these states with respect to the $\phi^+_{AB}$ is given by
\be\label{eq:Ext F}
F = \frac{1}{4}\big(1+\tr \, (\sigma_A(a) \sigma_B(b)) \big).
\ee
\bigskip

We now aim at finding a better map than the one temporarily fixed for Alice. 
So far, we used a straightforward translation from the CHSH case. However, we can design a better extraction map for Alice, exploiting the asymmetry of generalized CHSH tests.\\

In order to do so, we focus on the upper and lower lines of the frame corresponding to Bob settings $b=0$ and $b=\pi/2$. We label $F^\uparrow(a)$ the fidelity corresponding to the upper line of the frame and $F^\downarrow(a)$ the one of the lower line. Using \eq{Ext F} to express these fidelities, we find that $F^\uparrow(a)=F^\downarrow(a)$ for all Alice's measurement setting $a$. However, the generalized CHSH score of the state associated to positive eigenvalues of generalized Bell operators on the upper frame differs from such a score on the lower frame, as follows from the form of the function $f(b)$ in \tab{frame}. Indeed, these scores reads
\begin{equation}
    \begin{split}
        s^\uparrow = &\tr\,(\cB_\theta(a,0) (\cos(a)\sigma_h+\sin(a)\sigma_m)\otimes\sigma_z) = 2\sqrt{2}\cos(\theta), \\
        s^\downarrow = &\tr\,(\cB_\theta(a,\frac{\pi}{2})(\cos(a)\sigma_h-\sin(a)\sigma_m)\otimes\sigma_x) = 2\sqrt{2}\sin(\theta),
    \end{split}
\end{equation}
noting $s^\uparrow\geq s^\downarrow$ for all $a$, and for $\theta\in[0,\pi/4]$. \\

Since we are interested in the convex roof as a function of the generalized CHSH score (see Eq.~\eqref{eq:O}), an extraction map that yields the same fidelity for different Bell scores seems sub-optimal. Since the lower part of the frame results in a lower score, it is only the upper part which limits $\cF$. 
In particular, a map on Alice's side raising $F^\uparrow(a)$ could lead to a higher extractability. Since $F^\uparrow$ can be increased if $\sigma_A(a)$ gets closer to $\sigma_B(0)=\sigma_z,$ we append a rotation by an angle $\omega$ around $\sigma_y$ to Alice's extraction map. This will ultimately increase $F^\uparrow(a)$ at the price of lowering the fidelity $F^\downarrow(a)$, sending $\sigma_A(a)$ further away from $\sigma_B(\frac{\pi}{2})=\sigma_x$. \\

For some generalized CHSH tests with the parameter $\theta$ slightly lower that $\frac{\pi}{4},$ $s^\downarrow$ is small and plays essentially no role in limiting the final extractability. In that case, it is natural for Alice to try to increase $F^ \uparrow(a)$ as much as possible, ideally up to $\frac{1}{2}.$ However, for some settings $a$, it appears that $F^\uparrow(a)$ can not reach $\frac{1}{2}$, even when optimising the angle $\omega$ of the rotation. To overcome this limitation, we also tweak the direction of the dephasing associated to the map $\Lambda_A$. Concretely we introduce a parametric dephasing direction $\Gamma(d) = \cos(d)\sigma_h-\sin(d)\sigma_m$ allowing $F^ \uparrow(a)$ to take the value $\frac{1}{2}$ for any setting $a$. For simplicity, the strength of the dephasing is kept the same as before. \\

Lastly, we optimise the convex roof over the two parameters $\omega,d$, for a given setting $a$ and a given operator $\cB_\theta,$ that is, we choose $\omega,d$ such that 
\begin{equation}
    \label{eq:dirdeph_rot}
    \text{argmin}_{\omega,d} \left(\max\left\{ \frac{1-F_\uparrow(a)}{2\sqrt{2}-s^\uparrow},\frac{1-F_\downarrow(a)}{2\sqrt{2}-s^\downarrow}\right\} \right).
\end{equation} \\
Finally, Alice's map with the rotation around $\sigma_y$ and new dephasing direction $\Gamma(d)$ reads
\begin{equation}
    \nonumber
    \begin{split}
      \Lambda_A&(a)[\rho] = \\
      &U_a(\omega)\left(\frac{1+g(a)}{2}\id\rho\id + \frac{1-g(a)}{2}\Gamma(d)\rho\Gamma(d)\right)U_a(\omega)^\dagger 
      \end{split}
\end{equation}
with
\begin{equation}
      \omega,d = \text{argmin}_{\omega,d} \left(\max\left\{ \frac{1-F_\uparrow(a)}{2\sqrt{2}-s^\uparrow},\frac{1-F_\downarrow(a)}{2\sqrt{2}-s^\downarrow}\right\}\right)
\end{equation}
\section{Robustness of self-testing from generalized CHSH scores}

\begin{figure}[!ht]
\begin{center}
        \includegraphics[width=.48\textwidth]{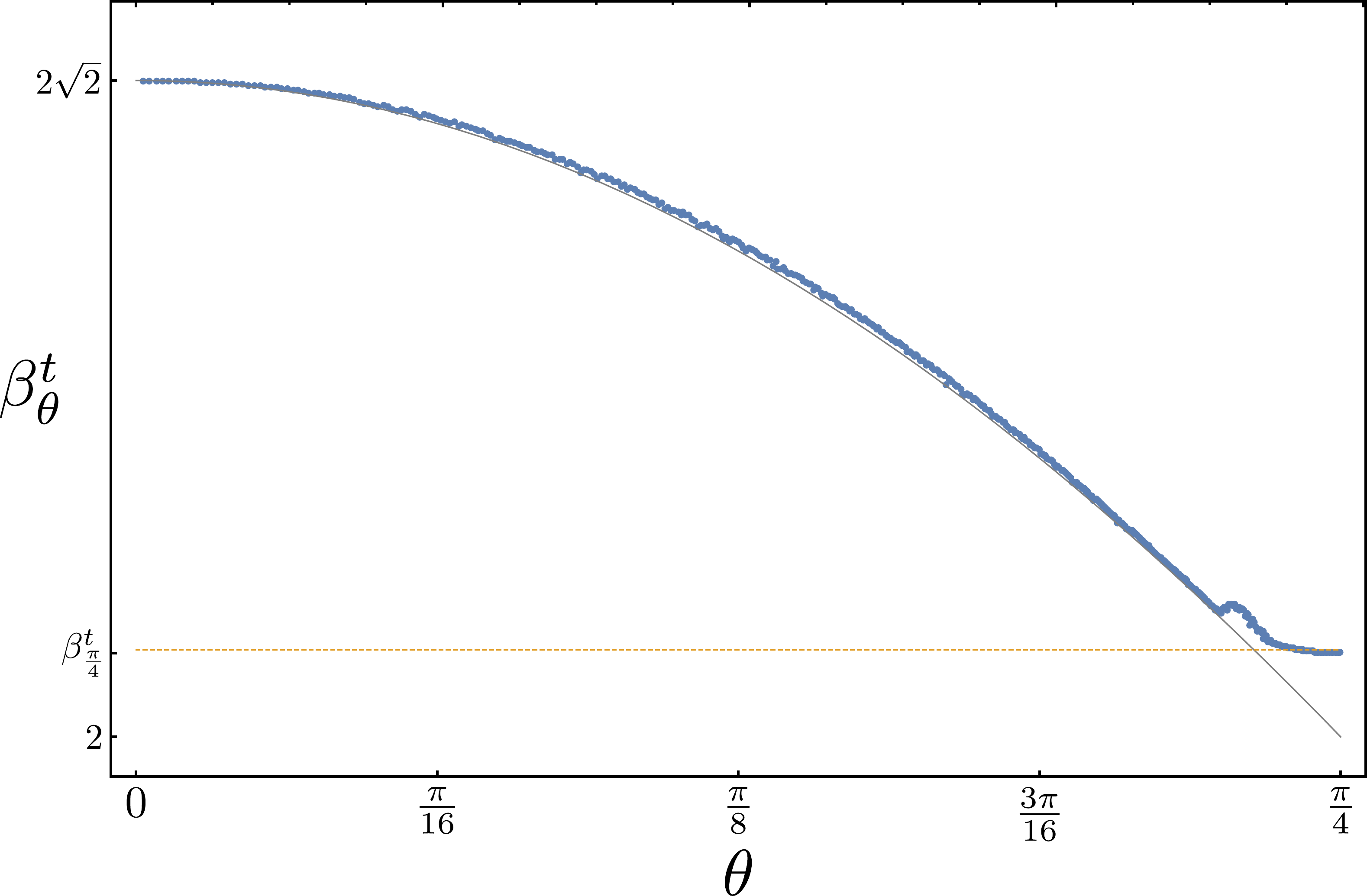}
        \caption{Minimum value $\beta_\theta^t$ of the expectation value of the generalized CHSH operators $\cB_\theta$ leading to a non-trivial self-testing bound $(\cF \geq 1/2)$ as a function of the parameter $\theta.$ For $\theta=\pi/4,$ $\cB_{\frac{\pi}{4}}$ corresponds to the CHSH operator and the trivial self-testing bound is $\beta_{\frac{\pi}{4}}^t \approx 2.11$ in agreement to the results of Refs.\cite{Kaniewski2016, Sekatski2018}.}
        \label{fig:betatrivial}
\end{center}        
\end{figure}

We first solve the two-qubit optimisation given in Eq. \eq{O} using the extraction maps proposed in the previous section and then deduce a lower bound $\cF$ on the extractability by taking the convex roof, as is detailed in the Appendix. The generalized CHSH score for which this lower bound reaches $\cF=1/2$ is referred to as the trivial score, denoted $\beta_\theta^t,$ that is, the trivial fidelity  $\cF=1/2$ is obtained for any violation satisfying $\beta_\theta\leq\beta_\theta^t$. For higher generalized CHSH scores, the bound increases linearly from $(\cF,\beta_\theta)=(1/2,\beta_\theta^t)$ to $(\cF,\beta_\theta)=(1,2\sqrt{2}),$ that is
\begin{equation}
    \label{eq:fid_xy}
    \cF_\theta(\beta_\theta) = \begin{cases}
        1/2\quad &\text{if}\, \beta_\theta\leq\beta_\theta^t,\\
        \frac{1}{2}\left(1+\frac{\beta_\theta-\beta_\theta^t}{2\sqrt{2}-\beta_\theta^t}\right) \quad &\text{otherwise.}
    \end{cases}
\end{equation} 
The values of the minimum generalized CHSH scores $\beta_\theta^t$ needed to get a non trivial self-testing bound are shown in Fig.~\ref{fig:betatrivial} as a function of $\theta$. The numerical values of $\beta_\theta^t$ can be found online on GitLab~\footnote{https://gitlab.com/plut0n/GenereralizedCHSH}.
\bigskip

 The previous formula can be used to self-test a source expected to produce two-qubit maximally entangled states from  generalized CHSH scores. This naturally rises the question of how to choose the generalized CHSH operator leading to the highest self-testing bound given an experiment in which values of $X$ and $Y$ are observed.

\section{Choosing the generalized CHSH test}

We consider the case where $X$ and $Y$ are given and clarify on the choice of the operator $\cB_\theta$ leading to the highest self-testing fidelity. \bigskip

Let us begin by discussing values of the pair $(X,Y)$ that are relevant for self-testing. First, we can assume $X,Y \geq 0$ without loss of generality since $X,Y \geq 0$ can always be obtained by relabelling the measurement outcomes of $A_1,$ $B_0$ and $B_1.$ We can also assume that $X-Y \geq 0$ without loss of generality because in case $X-Y < 0,$ we can relabel the inputs to switch $X$ and $Y$. The local bound of the CHSH operator $X+Y=2$ invites us to look at the region above the line connecting the deterministic local strategies $(X, Y) = (2, 0)$ and $(X, Y) = (0, 2),$ that is the region for which $X+Y\geq 2.$ Finally, the region achieved by quantum states and quantum measurements is bounded by $X^2+Y^2\leq 4,$ as shown in Ref.~\cite{Acin2012,Sekatski2020}. Hence, the values of the pair $(X,Y)$ that are relevant for self-testing can be reported in the positive quadrant with $X-Y \geq 0$, $X+Y\geq 2$, and $X^2+Y^2\leq 4.$ 
\bigskip

\begin{figure}[!ht]
\flushleft
\includegraphics[width=.47\textwidth]{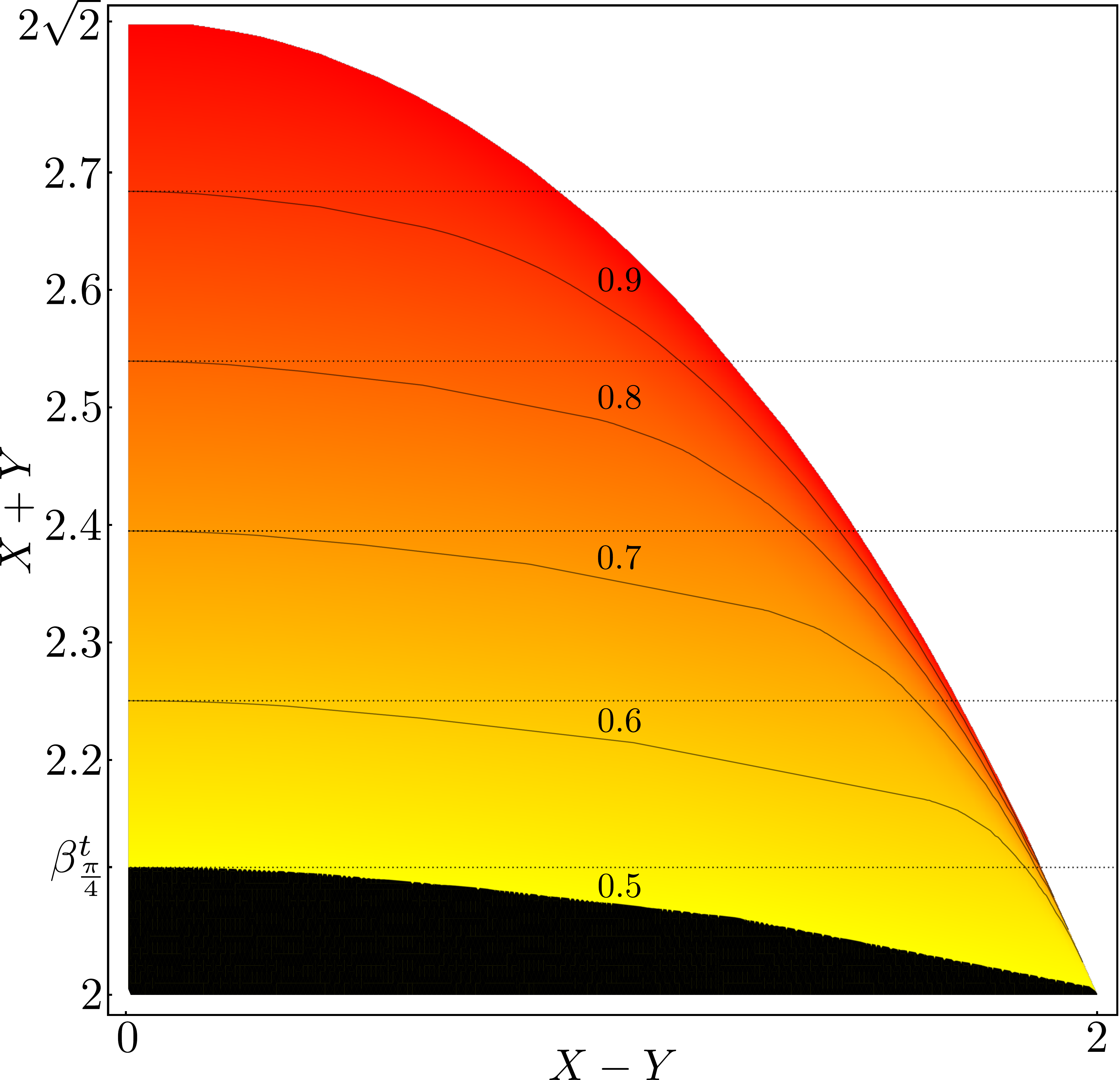}
\caption{Maximum of lower bounds on $\cF$ obtained from all all generalized CHSH operators $\cB_\theta.$ The black lines indicate values of $\cF$ by increment of $.1$ (label above the line). Lower bound on $\cF$ obtained from the CHSH score are given by the dashed lines (label below the line).}
\label{fig:fid}
\end{figure}

\begin{figure}[!ht]
\includegraphics[width=.47\textwidth]{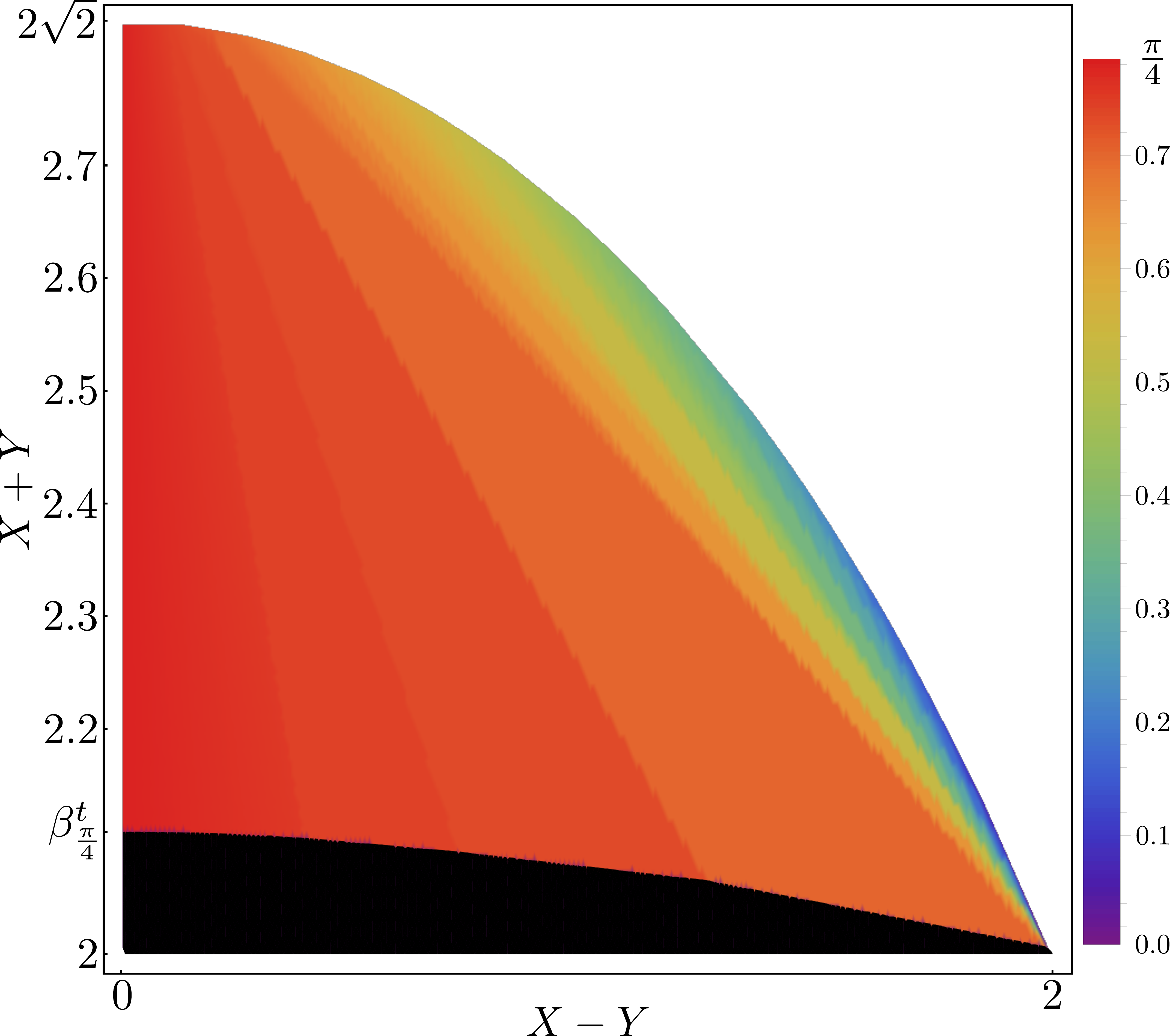}
\caption{Choice of generalized Bell operator $\cB_\theta$ leading to the highest lower bound on $\cF$
The black region corresponds to values of $\theta$ leading to trivial bounds on $\cF$.}
\label{fig:optimal_theta}
\end{figure}

We discretize this relevant region by meshing it with squares of side $\delta$, i.e.
\begin{align}
    |X-X'| \geq \delta,&\quad\forall X \neq X', \\
    |Y-Y'| \geq \delta,&\quad\forall Y \neq Y'.
\end{align}
Given a pair $(X,Y)$ on the mesh, we compute the values of generalized CHSH scores $\beta_\theta$ for 500 values of $\theta$ from Eq. ~\eqref{eq:btheta}. Note 
that it is sufficient to consider values of $\theta \in [0,\pi/4]$, since we can recover all cases $\theta\in [\pi/4,\pi/2]$ by permuting $X$ and $Y$.
For every value of $\theta$, we compute the lower bound on $\cF$ using the optimization procedure described above. The maximum values of these bounds over $\theta$ is represented in~\fig{fig:fid} while the value of $\theta$ leading to the corresponding maximum is shown in~\fig{fig:optimal_theta}. This gives a recipe for self-testing maximally entangled two-qubit states from generalized CHSH scores: from a pair $(X,Y),$ first deduce the most advantageous generalized CHSH operator from~\fig{fig:optimal_theta}. Then run the optimization described before to conclude on the self-testing bound, whose value is given in~\fig{fig:fid}.

Self-testing of a maximally entangled two-qubit state from the CHSH score only depends on the value of $X+Y.$ Known bounds on $\cF$ are reported as horizontal dashed lines in~\fig{fig:fid}. We see from the figure self-testing based on the generalized CHSH scores outperforms CHSH for all pairs of values $(X,Y)$ where $X \neq Y$. In particular, self-testing based on generalized CHSH tests is possible even for CHSH scores below $\approx 2.05$ where no self-testing statement can be obtained from the CHSH score alone~\cite{Valcarce2020}. Note that, in extreme cases where $X<<Y$, self-testing is possible even for CHSH scores arbitrary close to (but strictly larger than) the local bound.
\section{Conclusion}
The most robust self-test of two-qubit maximally entangled states was obtained so far from the CHSH score $\beta=\langle A_0(B_0+B_1) \rangle+\langle A_1(B_0-B_1) \rangle.$ Despite several previous attempts, it was not clear how the knowledge of individual values $\langle A_x B_y \rangle$ can be used to improve the robustness of such a self-test. Here, we derive a self-testing strategy based on the knowledge of the two individual terms $X=\langle A_0(B_0+B_1) \rangle$ and $Y=\langle A_1(B_0-B_1) \rangle$ appearing in the CHSH score. We show in particular that such a strategy is equivalent to considering a family of generalized CHSH operators, and go on to present an explicit recipe for choosing the appropriate operator given observed values for the pair $(X,Y)$. We prove that using the available information of $X$ and $Y$ improves the bound on the fidelity with respect to CHSH for all pairs $(X,Y)$, except the one where $X=Y$. Furthermore, self-testing statements can be obtained for some pair $(X,Y)$ where the strategy based on the CHSH test fails. Our results facilitate the experimental implementation of any device-independent certification based on the self-test of maximally entangled two-qubit states. 
\bigskip

\section{Acknowledgements}
We thank Jean-Daniel Bancal for fruitful discussions and comments. This work was supported by the Swiss National Science Foundation (SNSF) through the Grant PP00P2-179109 and the Region Ile-de-France in the framework of DIM SIRTEQ. 


\bibliography{references}{}
\bibliographystyle{ieeetr}

\bigskip
\appendix

\section*{Appendix}
\label{app:OptimizationCode}

A lower bound on the extractability $\cF$ can be obtained as the convex roof of the two-qubit optimisation $\cO(\beta_\theta')$ defined in \eq{O}, over all generalized CHSH score $\beta_\theta'$. In this appendix, we give more details on how the optimization is performed.\\

For a given score $\beta_\theta'$, we minimize the fidelity with respect to $\phi^+_{AB}$ over all two-qubit states and over all angles of measurement achieving a violation of at least $\beta_\theta'.$ Such an optimisation is formally expressed as

\begin{equation}
    \label{eq:opt_instance}
      \begin{array}{@{}>{\displaystyle}l@{}}
          \min_{a,b} \left( \min_{\rho_{AB}^{\text{qubit}}} \left(\tr ((\Lambda_A(a) \otimes \Lambda_B(b))[\rho_{AB}^{\text{qubit}}]\, \phi^+_{AB}) \right)\right) \\
          \textnormal{subject to:} \\
          \quad \tr (\cB_\theta(a,b)\rho_{AB}^{\text{qubit}}) \geq \beta_\theta', \\
          \quad \rho_{AB}^{\text{qubit}} \succ 0, \\
          \quad \tr (\rho_{AB}^{\text{qubit}}) = 1, \\
          \quad (\rho_{AB}^{\text{qubit}})^\dagger = \rho_{AB}^{\text{qubit}}.
      \end{array}
\end{equation}

with $\rho_{AB}^{\text{qubit}}\in\mathbb{C}^2\otimes\mathbb{C}^2$ and $(a,b)\in\left[0,\frac{\pi}{2}\right]$.\\

As a reminder, for a given operator $\cB_\theta$, Alice's map $\Lambda_A(a)$ also includes an inner optimisation (see \eq{dirdeph_rot}) over two scalars $\omega$ and $d$. However, in our case this optimization is independent of the minimization above, and is executed beforehand, fixing the maps $\Lambda_A(a)$ in \eq{opt_instance}. 
\bigskip

In this Appendix, we will explore each optimization step independently. We start by focusing on the parametrisation of Alice's map. Then, we detail the semi-definite programming optimisation over two-qubit states. We further discuss the optimisation of Alice and Bob settings. Finally, we show how the convex roof is obtained to get the final lower bound on $\cF$. \bigskip

\subsection{Inner optimization of Alice's map parameters}
\label{app:AliceParamOpt}

For a generalized CHSH operator $\cB_\theta$ and Alice's measurement setting $a$, we need to find a good choice of Alice's map parameters. However, the best choice of parameters is not required since sub-optimal maps only result in underestimating the self-testing capabilities of generalized CHSH self-test. \\
As explained in the main text, Alice's map is parametrized from the result of the following optimization 
\begin{equation}
    \label{eq:dirdeph_rot2}
    \small
    \begin{split}
    \Lambda_A(a;\theta) =    &\, U_a(\omega)\left(\frac{1+g(a)}{2}\id\rho\id \right.\\
    &\quad \left.+ \frac{1-g(a)}{2}\Gamma(d)\rho\Gamma(d)\right)U_a(\omega)^\dagger   \\
    \text{with}&\\
    (\omega,d) =& \text{argmin}_{\omega,d} \bigg(\\ 
    & \left.\max\left\{ \frac{1-F_\uparrow(a,\omega,d)}{2\sqrt{2}-s^\uparrow},\frac{1-F_\downarrow(a,\omega,d)}{2\sqrt{2}-s^\downarrow}\right\} \right).
    \end{split}
\end{equation}
The quantities $F^\uparrow,F^\downarrow$ are defined using \eq{Ext F} and read
\begin{equation}
    \begin{split}
        F_\uparrow(a,\omega,d) &= \frac{1}{4}\left(1+\sigma_A'(a,\omega,d)\sigma_B(0)\right) \\
        F_\downarrow(a,\omega,d) &= \frac{1}{4}\left(1+\sigma_A'(a,\omega,d)\sigma_B(\frac{\pi}{2})\right).
    \end{split}
\end{equation}
$\sigma_A'(a,\omega,d)$ results from the action of Alice's map on the states associated to the positive eigenvalues of the Bell operator $\cB_\theta(a,b)$. Formally
\begin{equation}
    \small
    \sigma_A'(a,\omega,d) = \underbrace{{}R(\omega-d)\left(\begin{array}{cc}
        1 & 0 \\
        0 & g(a)
    \end{array}\right)
    R(d)}_{\Phi_A(a,d,\omega)}\colvec{\cos(a)\sigma_h \\ \sin(a)\sigma_m}.
\end{equation}
where $R(\cdot)$ is a SO(2) rotation. With the above expression, fidelities on the frame simplifies to 
\begin{equation}
\small
\label{eq:fidframe}
    \begin{split}
        F^\uparrow(a,\omega,d) &= \frac{1}{4}\left(1+\colvec{\frac{1}{\sqrt{2}} \\ \frac{1}{\sqrt{2}}}\Phi_A(a,d,\omega)\colvec{\cos(a)\\\sin(a)}\right), \\
        F^\downarrow(a,\omega,d) &= \frac{1}{4}\left(1+\colvec{\frac{1}{\sqrt{2}} \\ -\frac{1}{\sqrt{2}}}\Phi_A(a,d,\omega)\colvec{\cos(a)\\-\sin(a)}\right). \\
    \end{split}
\end{equation} \\
\eq{dirdeph_rot2} is solved using the expression \eq{fidframe} for the fidelities. 
\bigskip

For a given Bell operator $\cB_\theta$, we have to run the optimisation given in \eq{dirdeph_rot2} for each value of $a$. Since we have to optimize over Alice's measurement setting to get $\cF$, running the this optimisation can be resource-heavy. \\
To shorten each of these optimisations, we start by creating a discrete set of Alice measurement setting $a$. We choose a set of 100 settings ${\mathbb a}=\{a_n\}_{n=1}^{100}$ with a step $\delta = \frac{\pi/4}{100}$ so that $a_n=a_{n-1}+\delta$. Then, for each element of the set, $a_i$, we run the optimisation given in \eq{dirdeph_rot2} to obtain the parameters $\omega_i$ and $d_i$. In practice, this optimisation is run multiple times, with different starting points at each run. This will increase our confidence in the result. \\
The set of results obtained this way, is used as a starting point for the optimization of values of $a$ that do not belong to the grid. This is, for $a\notin{\mathbb a}$ we run an optimisation, e.g. using a Limited-memory BFGS method, with initial guess $\omega_i,d_i$, the parameters obtained for $a_i\approx a$, where $a_i\in{\mathbb a}$. \bigskip

\subsection{Semidefinite programming for optimising over two-qubit states}
\label{app:sdp2qubit}

Finding the worst-case fidelity over all two-qubit states is an optimization over Hermitian matrices $\rho_{AB}^{\text{qubit}}$ that are valid density matrices, i.e positive semi-definite with unit trace. Furthermore, the objective function 
\begin{equation}
    \cO = \tr ((\Lambda_A(a,\omega,d)\otimes\Lambda_B(b))[\rho_{AB}^{\text{qubit}}]\, \phi^+)
\end{equation} 
depends linearly on $\rho_{AB}^{\text{qubit}}$, and the constraints are affine expressions of this variable. Such a problem can be solved with semidefinite programming (SDP). Interestingly, SDP is a certifiable method of optimisation. This is, either the optimisation is \textit{infeasible} or the optimisation results is tight -- up to the numerical tolerance, in a case of weak duality. \bigskip

\subsection{Minimization over the angles and confidence}
\label{app:minangleconf}

Minimizing over the choices of measurement, $a,b$, is a non-convex optimization over two scalars in a closed domain $(a,b) \in ([0,\frac{\pi}{2}] \times [0,\frac{\pi}{2}])$. In practice, this optimisation is performed using the Scipy implementation of the L-BFGS algorithm. \\

\paragraph{Infeasible SDP --} Some measurement choices result in an infeasible SDP, since for some  measurement angles $a,b$, there exist no state $\rho_{AB}^{\text{qubit}}$ satisfying the generalized CHSH score constraint $\tr (\cB_\theta(a,b)\rho_{AB}^{\text{qubit}}) = \beta_\theta'$. \\

When this happens, in order to help the optimisation converge to a feasible solution, we guide the angles $(a,b)$ towards the optimal measurement setting for which we know any Bell score $\beta_\theta'\leq 2\sqrt{2}$ can be attained and, thus, the SDP is feasible. 
To do so we introduce the function
\begin{equation}
    V = 1 + \left|\left|\colvec{a-\frac{\pi}{4} \\ b-\theta}\right|\right|^2\,.
\end{equation}
It takes the minimum value $1$ for $(a,b)=(\frac{\pi}{4},\theta)$ and increases with the distance to the optimal choice of measurement. \\
At each iterative step of the $(a,b)$ minimization, if the SDP is found infeasible, the objective function of the optimisation is replaced by $V$ for this step, i.e.
\begin{equation}
    \small
    \begin{cases}
    \min_{\rho_{AB}^\text{qubit}}\cO, &\text{if}\; \exists\, \rho_{AB}^{\text{qubit}}\; \text{s.t}\;\tr (\cB_\theta(a,b)\rho_{AB}^{\text{qubit}}) = \beta_\theta' \\
    V, & \text{otherwise}. 
    \end{cases}
\end{equation}
Therefore, at a given optimization iteration if the SDP is infeasible the minimisation over $a,b$ will follow the gradient $V$, exploring parameters toward the optimal choice of measurement angles. \bigskip

\paragraph{Entrusted results of minimisation --} From the fact that the minimization comes with no certificate, it is possible that we do not find the angles $(a,b)$ attaining the minimal fidelity. For a given generalized CHSH score, this could result in a higher singlet fidelity and, thus, an overestimation of $\cF$. To build trust in the numeric results we implement several routines. \bigskip

A first routine is made to improve the confidence in the minimisation outcome. This is, the minimisation over the choices of measurement is run a fixed number of time $n$ for a given violation. \bigskip

Then, for a given generalized CHSH score, we make sure that the minimum fidelity found is lower than the one obtained for an higher score.
A pseudo-code of these routines can be found in \alg{alg:min}.

\begin{algorithm}[!hb]
    \SetAlgoLined
     \text{Input} $\theta$,$k$ \\
     $\beta_\theta^k \gets $~\eq{betak} \\
     $m_F(\beta_\theta^k) \gets \infty$ \\
     \For{$i\gets0$ \KwTo $n$}{
        $F_i(\beta_\theta^k)$ $\gets$ instance of \eq{opt_instance} for $\beta_\theta^k$\\
       \If{$m_f \geq F_i(\beta_\theta')$}{
            $m_F(\beta_\theta') \gets F_i(\beta_\theta^k)$\\
        }
    }
    \If{$m_F(\beta_\theta^k) < m_F(\beta_\theta^{k-1})$}{
        \alg{alg:min}($\theta$,$\beta_\theta^{k}$)
    }
    \text{Return }$m_F(\beta_\theta^k)$ 
\caption{Minimization of the fidelity for $\cB_\theta$}
\label{alg:min}
\end{algorithm}
\bigskip

\subsection{Numerical values of trivial scores $\beta_\theta^t$}
\label{app:findbetatineq}

From the optimisation given in \eq{opt_instance}, we have the minimum fidelity over all two-qubit states for a fixed Bell operator $\cB_\theta$ and generalized CHSH score $\beta_\theta'$. To get $\cF$, we need the convex roof of this minimum fidelity over all scores $\beta_\theta'\in[-2\sqrt{2},2\sqrt{2}]$. It is however unnecessary to explore this whole range of scores. Indeed, when computing the minimum fidelity over all scores for different Bell operator $\cB_\theta$, we found that $\beta_\theta*$, the score from which the convex roof is taken from, is close to the local bound and the only point of inflection. Using this ansatz, we thus start by solving \alg{alg:min} on $\beta_\theta'=\beta_\theta^L$. We then decrease the generalized CHSH score by a step $\varkappa$ defined as
\begin{equation}
    \label{eq:varkappa}
    \varkappa = \left(\frac{\theta}{\pi/4}\right)^2 \kappa
\end{equation}
where, in practice, we set $\kappa=0.025$. The resulting score at the $k$-step is given by
\begin{equation}
    \label{eq:betak}
    \beta_\theta^k = \beta_\theta^L-k\varkappa.
\end{equation}
\bigskip

For each score $\beta_\theta^k$ we run \alg{alg:min}. Then, we compute the slope between the  minimum fidelity obtained for $\beta_\theta^k$, $m_F(\beta_\theta^k)$ and the quantum bound with fidelity 1. The slope reads
\begin{equation}
    \alpha(\beta_\theta') = \frac{1-F(\beta_\theta')}{2\sqrt{2}-\beta_\theta'}.
\end{equation}
At each step $k$ we check if this slope, $\alpha(\beta_\theta^k)$, is greater than the one obtain at the previous step, $\alpha(\beta_\theta^{k-1})$. If this condition is not satisfied, it means than a point of inflection was reach at the step $k-1$. If so, we label the pair score-minimum fidelity of the inflexion point as $(\beta_\theta^*,m_F^*)$. \bigskip

Finally, the trivial generalized CHSH score $\beta_\theta^t$ is obtained from the intersection between the convex roof taken at the inflection point and the line $F=\frac{1}{2}$ using
\begin{equation}
    \beta_\theta^t = \frac{0.5-m_F(\beta_\theta^*)}{\alpha(\beta_\theta^*)}+\beta_\theta^*.
\end{equation}
\end{document}